\documentclass[prb, aps, longbibliography, showpacs, superscriptaddress, twocolumn]{revtex4-2}

\usepackage{color}
\usepackage[usenames,dvipsnames]{xcolor}
\usepackage{amsmath,amsthm,amssymb}
\usepackage{graphicx}
\usepackage{float}
\usepackage{epsfig}
\usepackage{bm}
\usepackage{mathrsfs}
\usepackage{multirow}
\usepackage[all]{xy}
\usepackage{pbox}
\usepackage{verbatim}
\usepackage{braket}
\usepackage{mathtools}
\usepackage{bm}
\usepackage{tikz}
 
\usepackage{mathtools}
\usepackage{tabstackengine}
\usepackage{enumerate}   
\usepackage{wasysym}
\usepackage{nicematrix}

\stackMath
\DeclareMathOperator{\Tr}{Tr}

\newcommand{\er}[1]{Eq.~\eqref{#1}}

\usepackage{hyperref}
\hypersetup{
     colorlinks=true,
     linkcolor=magenta,
     filecolor=blue,
     citecolor=blue,      
     urlcolor=cyan,
     }

\newcommand{\beq}{\begin{equation}}
\newcommand{\eeq}{\end{equation}}

\newcommand{\D}{\mathcal D}

\begin{document}  

\title{Non-thermal eigenstates and slow relaxation in quantum Fredkin spin chains}

\author{Luke Causer}
\affiliation{School of Physics and Astronomy, University of Nottingham, Nottingham, NG7 2RD, UK}
\affiliation{Centre for the Mathematics and Theoretical Physics of Quantum Non-Equilibrium Systems,
University of Nottingham, Nottingham, NG7 2RD, UK}
\author{Mari Carmen Ba\~nuls}
\affiliation{Max-Planck-Institut f\"ur Quantenoptik, Hans-Kopfermann-Str.\ 1, D-85748 Garching, Germany}
\affiliation{Munich Center for Quantum Science and Technology (MCQST), Schellingstr.\ 4, D-80799 M\"unchen}
\author{Juan P. Garrahan}
\affiliation{School of Physics and Astronomy, University of Nottingham, Nottingham, NG7 2RD, UK}
\affiliation{Centre for the Mathematics and Theoretical Physics of Quantum Non-Equilibrium Systems,
University of Nottingham, Nottingham, NG7 2RD, UK}

\begin{abstract}
    We study the dynamics and thermalization of the Fredkin spin chain, 
    a system with local three-body interactions, particle conservation and explicit kinetic constraints.
    We consider deformations away from its stochastic point in order to tune between regimes where kinetic energy dominates and those where potential energy does. By means of exact diagonalisation, perturbation theory and variational matrix product states, 
    we show there is a sudden change of behaviour in the dynamics that occurs, from one of fast thermalization to one of slow metastable (prethermal) dynamics near the stochastic point.
    This change in relaxation is connected to the emergence of additional kinetic constraints which lead to the fragmentation of Hilbert space in the limit of a large potential energy.
    We also show that this change can lead to thermalization being evaded for special initial conditions
    due to non-thermal eigenstates (akin to quantum many-body scars). We provide clear evidence for the existence of these non-thermal states for large system sizes even when far from the large-potential-energy limit, and explain their connection to the emergent kinetic constraints.
\end{abstract}

\maketitle

\section{Introduction}\label{sec: introduction}
Recent years have seen many developments in the understanding of the dynamics and thermalization of quantum many-body systems.
The eigenstate thermalization hypothesis (ETH)~\cite{Deutsch1991,Srednicki1994} implies that, for generic models, the long-time local properties of large isolated quantum many-body systems are determined entirely by the energy density of the system: the conditions of the initial state are lost, and the expectation values of local observables are described by the thermal ensemble at a temperature given by the conserved energy (for reviews, see e.g. Refs.~\cite{DAlessio2016,Deutsch2018, Mori2018}).
While the ETH has been extensively verified both numerically (e.g. Refs.~\cite{Rigol2008, Beugeling2014,Beugeling2015, Dymarsky2018, Sugimoto2022}) and experimentally (e.g. Refs.~\cite{Trotzky2012, Kaufman2016, Neill2016, Govinda2016, Wang2022}) for a vast number of scenarios, there has been a great deal of interest in understanding systems which do not obey it.

A prime example of systems which violate the ETH are integrable systems (for reviews, see e.g. Refs.~\cite{Doikou2010, Bertini2021}).
Such systems have an extensive number of conserved quantities, allowing them to retain key information from initial conditions and avoid thermalization \cite{Mori2018} (converging instead to the so-called generalised Gibbs ensemble \cite{rigol2007relaxation,vidmar2016generalized}). A second example of non-ergodicity is thought to be that of many-body localisation (MBL) \cite{Basko2006, Oganesyan2007, nandkishore2015many-body,abanin2019colloquium}, where the combination of interactions and strong quenched disorder leads to the proliferation of emergent conserved quantities (although it is still debated whether MBL is truly stable or only metastable \cite{suntajs2020quantum,kiefer-emmanouilidis2021slow,morningstar2022avalanches,sels2022bath-induced}). 

Given the above, there has been a great interest in understanding what conditions can yield non-thermalising dynamics beyond the paradigm of integrability, in particular in the absence of quenched disorder. 
There are a number of frameworks now understood to lead to non-thermal behaviour even at long times. Examples include translationally invariant systems with boundary-localised almost-conserved operators, or ``strong zero modes'' \cite{Fendley2012, Sarma2015, Fendley2016, Kemp2017, Else2017, Vasiloiu2019}, systems with kinetic constraints \cite{Horssen2015, Hickey2016, Feldmeier2019, Pancotti2020, Bertini2023}, 
systems in tilted potentials \cite{nieuwenburg2019from,schulz2019stark,kloss2023absence}, 
lattice gauge systems \cite{smith2017disorder-free,brenes2018many-body,karpov2021disorder-free,Sau2024},
and (dissipative) non-Hermitian quantum Hamiltonians displaying ``skin effects'' \cite{Cipolloni2023,znidaric2022solvable}.
A promising avenue of research is the recently discovered area of {\it quantum many-body scars} (QMBS) \cite{Turner2018, Turner2018b} (for reviews see Refs.~\cite{Serbyn2021, Moudgalya2022, Papic2022, Chandran2023}). This is the name given to non-thermal eigenstates in an otherwise thermalising Hamiltonian. Originally discovered \cite{Turner2018} in the PXP constrained model \cite{fendley2004competing,lesanovsky2011many-body} to explain the non-thermal behaviour observed in cold atom experiments \cite{Bernien2017, Bluvstein2021controlling, Su2023observation}, there are now many systems known to host QMBS, see for example Refs.~\cite{Moudgalya2018, Lin2019, Schecter2019, Choi2019, Zhao2020, Langlett2021, Francica2023, Zhao20121, Richter2022}.

In this paper, we study quantum Fredkin spin chains \cite{Salberger2016} with an additional parameter to deform the Hamiltonian away from its stochastic point in order to explore slow thermalisation and the existence of a family of non-thermal eigenstates. This is similar in spirit to the study in Ref.~\cite{Pancotti2020} of the kinetically constrained quantum East model \cite{Horssen2015}, which showed the existence of a large number of non-thermal eigenstates responsible for its slow relaxation dynamics. Like in the quantum East model, 
here we find that at the stochastic point, there is a distinct change in the dynamical behaviour from fast relaxation to slow thermalisation.
Furthermore, we show that in this model there exist non-thermal eigenstates reminiscent of QMBS.
We explain this behaviour through the emergence of a {\it folded} model \cite{Zadnik2021, Zadnik2021b}, that is, a quantum dynamics that is more constrained than what is explicitly stated by the dynamical rules of the Hamiltonian, and which is responsible for the intermediate time dynamics. We note that, while having some similarities, the non-thermal eigenstates that we find here for the Fredkin model are distinct from those of Refs.~\cite{Langlett2021,Francica2023} which consider generalised Fredkin chains.

The paper is organised as follows.
In Sec.~\ref{sec: system} we introduce the model and its Hamiltonian, and describe its symmetries and its known ground-state phase diagram \cite{Sugino2018, Causer2022}.
Section \ref{sec: folded} explains the emergence of the folded model in the large potential energy limit: this is a more constrained version of the original Hamiltonian that helps explain the behaviour of the original model by means of perturbation theory.
We numerically demonstrate nonergodic properties of the system in Sec.~\ref{sec: relaxation}: while the level spacing statistics suggest that the system is ergodic overall, we are able to find interesting slow and heterogeneous dynamics evidenced through autocorrelation functions and growth of entanglement entropy. In Sec.~\ref{sec: spectrum}, we show that this is a consequence of non-thermal eigenstates: we are able to find these eigenstates exactly for small system sizes using exact diagonalisation (ED), and 
provide convincing evidence for their existence in larger systems by approximating them using variational matrix product states (MPS) and perturbation theory. We conclude in Sec.~\ref{sec: conclusions}, where we offer an outlook on our results.

\section{Model} 
\label{sec: system}
We consider a one-dimensional lattice of $N$ spin-$1/2$ particles, each with local basis states $\ket{\newmoon_{i}}$ and $\ket{\fullmoon_{i}}$, where $i$ denotes the lattice site.
We will refer to the basis states as spin up and spin down respectively, or alternatively, particles and holes.
The system evolves under a Hamiltonian with local three-body interactions,
\begin{multline}
    \hat{H}_{c, s} = - \sum_{i=1}^{N-1}
    \hat{f}_{i}
    \Big[e^{-s} \sqrt{c(1-c)} \big( \hat{S}_{i}^{+}\hat{S}_{i+1}^{-}
    + \hat{S}_{i}^{-}\hat{S}_{i+1}^{+} \big) 
    \\ 
    - c\hat{v}_{i}\hat{n}_{i+1} 
    - (1-c)\hat{n}_{i}\hat{v}_{i+1} \Big],
    \label{H}
\end{multline}
with the Pauli ladder operators acting on site $i$, $\hat{S}^{+}_{i} = \ket{\newmoon_{i}}\bra{\fullmoon_{i}}$ and $\hat{S}^{-}_{i} = \ket{\fullmoon_{i}}\bra{\newmoon_{i}}$, and occupation operators $\hat{n}_{i} = \ket{\newmoon_{i}}\bra{\newmoon_{i}}$ and $\hat{v}_{i} = \ket{\fullmoon_{i}}\bra{\fullmoon_{i}}$.
The operator $\hat{f}_{i} = \hat{n}_{i-1} + \hat{v}_{i+2}$ is the Fredkin {\it kinetic constraint}.
This constraint only allows the terms inside the brackets of \er{H} to be ``activated'' if either the neighbour to the left of the pair is spin up, or if the neighbour to the right is spin down.
We allow for $c \in (0, 1)$ and $s \in (-\infty, \infty)$.
and consider open boundary conditions (OBC) with $\hat{v}_{0} = 1$ and $\hat{n}_{N+1} = 1$ fixed. 

The Hamiltonian $\hat{H}_{c, s}$ for $c = 1/2$ and $s = 0$ was introduced in Ref.~\cite{Salberger2016}, and was later generalised to $c \neq 1/2$ in Refs.~\cite{Salberger2017,Zhang2017, Sugino2018}. In Ref.~\cite{Causer2022}, the operator \er{H} was considered in the context of classical stochastic dynamics: 
with opposite sign,
the Hamiltonian \er{H} 
is explicitly a stochastic generator (of continuous-time Markov chains) for $c=1/2$ and $s=0$, while for $c \neq 1/2$ and $s = 0$, \er{H} is also stochastic but only after a similarity transformation (see Ref.~\cite{Causer2022} for details). In the stochastic context, the introduction of the $e^{-s}$ prefactor to the kinetic terms makes \er{H}  
equivalent to a {\it tilted generator} --- a classical stochastic dynamics which does not conserve probability and encodes the statistics of hopping events --- which allows one to study the fluctuations of the dynamics at $s = 0$. For the quantum case, the introduction of this parameter allows us to control the relative strength of the kinetic terms compared to the potential energy in the Hamiltonian. The generalisation of tilted generators to unitary quantum dynamics was recently considered in Refs.~\cite{Horssen2015,lan2018quantum,Pancotti2020} for other kinetically constrained models.

\subsection{Symmetries}

\begin{figure}[t]
    \centering
    \includegraphics[width=0.9\linewidth]{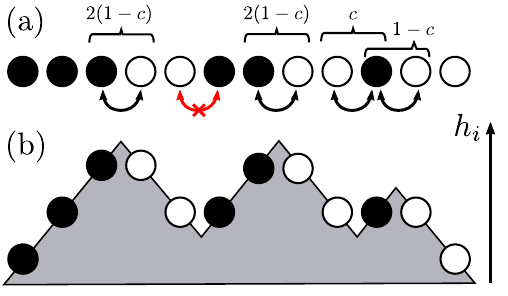}
    \caption{\textbf{Fredkin spin chain.} 
    (a) An example configuration from the largest subspace $\mathcal{D}$ for $N = 12$ sites. The (black) arrows show particle hops allowed by the constraints in the kinetic part of \er{H}. The (red) crossed arrow is a transition not allowed by the constraints.
    Above the configuration, we indicate the contributions to the energy from the diagonal operators in \er{H}.
    (b) The same configuration in its height field representation: a particle (hole) corresponds to a step up (down) in height. The (shaded) area 
    $\hat{A}$
    of the height field is an order parameter for the system.
    }
    \label{fig: model}
\end{figure}

The Fredkin model has a $U(1)$ symmetry which conserves the total magnetisation, $\hat{Z}_{\rm tot} = \sum_{i=1}^{N} \hat{Z}_{i}$, where $\hat{Z}_{i} = 2\hat{n}_{i}-1$ is the Pauli-$z$ operator acting on site $i$ \cite{Salberger2016}.
Furthermore, the kinetic constraint reduces the fixed magnetisation sectors into sub-sectors which can be understood in terms of random walk excursions and Catalan combinatorics \cite{Salberger2016}. The largest of these sub-sectors, $\mathcal{D}$, occurs for half-filling ($\hat{Z}_{\rm tot} = 0$) and corresponds to all the configurations which have at least as many spin ups as spin downs when counted from left-to-right in the lattice 
\cite{Salberger2016,Causer2022}. 
An example configuration in this sector is shown in Fig.~\ref{fig: model}(a). 
The figure shows the transitions that are allowed by the kinetic part of \er{H}. Above the configuration in Fig.~\ref{fig: model}(a) are the local contributions to the potential energy of \er{H}. 

It is possible to represent each configuration of $\mathcal{D}$ by a random walker excursion \cite{Majumdar2005} if one considers each lattice site as a step in time, where a particle moves the walker a step in the positive direction, and a hole moves it in the negative direction. This results in an alternative representation of configurations in terms of a {\it height field} $\hat{h}_{i} = \sum_{j=1}^{i} \hat{Z}_{j}$, see Fig.~\ref{fig: model}(b). In what follows we consider the case of half-filling ($Z_{\rm tot} = 0$) and the allowed configurations in that sector have non-negative height, $h_{i} \geq 0$, with the terminating condition $h_{N} = \hat{Z}_{\rm tot} = 0$. 
This set of classical basis states is therefore equivalent to Dyck paths, and this subspace has a dimension equal to the Catalan number $C_{N/2}$ \cite{Salberger2016}.
A natural observable to quantify a configuration is the {\it area} under the height field,
\beq
    \hat{A} = \sum_{i = 1}^{N} \hat{h}_{i} = \sum_{i = 1}^{N} (N + 1 - i)\hat{Z}_{i}.
    \label{eq:area}
\eeq
Notice that the bunching of particles produces a large area, while if they spread out the area becomes smaller.

The final symmetry is {\it charge parity} (CP).
The first aspect to this is a global spin-flip, $\mathcal{C} \ket{\fullmoon_{j} / \newmoon_{j}} = \ket{\newmoon_{j} / \fullmoon_{j}}$, and the second is spatially reflecting the lattice, $\mathcal{P} \ket{\fullmoon_{j} / \newmoon_{j}} = \ket{\fullmoon_{N+1-j} / \newmoon_{N+1-j}}$.
Together, this can be written as $\mathcal{CP} \ket{\fullmoon_{j} / \newmoon_{j}} = \ket{\newmoon_{N+1-j} / \fullmoon_{N+1-j}}$.
Note that this acts on all lattice sites simultaneously, and it follows that $[\hat{H}, \mathcal{CP}] = 0$.
Each of the sectors previously described (except for the frozen irreducible configurations $\fullmoon\cdots \fullmoon$ and $\newmoon\cdots\newmoon$) will contain two CP sectors, with ${\rm CP} = \pm 1$.
Note that there exist computational basis states $x$ that are invariant under the action of CP, and thus can only belong to the ${\rm CP} = +1$ sector.
Among these are the states $\ket{P_{j}} \in \mathcal{D}$,
\begin{multline}
    \ket{P_{j}} = \underbrace{\ket{\newmoon\cdots \newmoon}}_{N/2 + 1 - j}
     \otimes \underbrace{\ket{\fullmoon\newmoon \cdots \fullmoon\newmoon}}_{2j-2}
     \otimes \underbrace{\ket{\fullmoon \cdots \fullmoon}}_{N/2+1-j},
    \label{Pj}
\end{multline}
for $j = 1, \dots, N/2$, where the numbers underneath indicate the number of lattice sites.
These product states will play a significant role in understanding the non-thermal eigenstates states in Sec.~\ref{sec: spectrum}.

\section{Folded model} \label{sec: folded}

\begin{figure}[t]
    \centering
    \includegraphics[width=\linewidth]{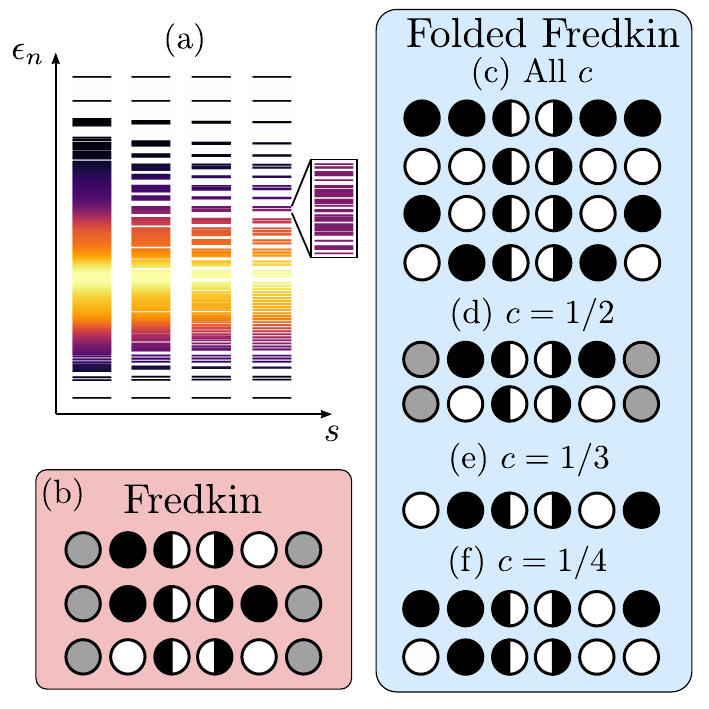}
    \caption{\textbf{Folded Fredkin model.}
    (a) The energy spectra of $\hat{H}_{c, s}$ (normalised between $0$ and $1$) for increasing $s$. The colours indicate the density of states in the local neighbourhood. As the value of $s$ is increased, the spectrum becomes sparser. The data is for $c = 0.7$ and $N = 18$ from ED. 
    (b) In the original Fredkin model, hops can occur between the middle sites (shown as half-filled circles) only for the cases shown, while the sites in grey do not participate in the constraint.
    (c) Transitions in the folded Fredkin model for any value of $c$ are allowed if both the nearest neighbours and next nearest neighbours to the central sites obey an XNOR constraint (i.e., they are equal). 
    (d) In the folded Fredkin model at $c = 1/2$ 
    the constraint reduces to an XNOR on the nearest neighbour pair, with the next nearest neighbours playing no role. 
    (e) For the special case of $c = 1/3$ there is an extra allowed transition on top of those for $c = 1/2$ as shown. 
    (f) Something similar occurs for the special case of $c = 1/4$, with two extra transitions allowed beyond those of $c = 1/2$.}
    \label{fig: folded}
\end{figure}

We now consider the Hamiltonian $\hat{H}_{c, s}$ in the limit $s \to \infty$.
It is first instructive to study the energy spectrum
of $\hat{H}_{c, s}$ restricted to $\mathcal{D}$, that is, 
$\{ E_n : E_n \leq E_{n+1} \}$, where $\hat{H}_{c, s}\ket{E_{n}} = E_{n}\ket{E_{n}}$ for all $n=1, 2, \cdots, {\rm dim}(\D)$. For convenience we normalise this spectrum between zero and one,
\beq
    \epsilon_{n} = \frac{E_{n} - E_{\rm min}}{E_{\rm max} - E_{\rm min}} ,
    \label{energy_density}
\eeq
where $E_{\rm min}$ ($E_{\rm max}$) are the minimum (maximum) energies in the ensemble. 
Figure~\ref{fig: folded}(a) shows the normalised spectrum for increasing values of $s \geq 0$ at $c = 0.7$ and $N = 18$. The colour of the lines indicates the density of states in the local neighbourhood with brighter indicating a higher density. Since the Hamiltonian has the form $\hat{H} = e^{-s}\hat{T} + \hat{V}$, increasing the value of $s$ increases the relative strength of the potential energy $\hat{V}$ terms to the kinetic energy $\hat{T}$. This causes the spectrum to look more sparse with increasing $s$, forming distinct bands around the discrete eigenvalues of $\hat{V}$ (noting that there is still repulsion of levels within the bands due to $\hat{T}$). 

This behaviour can be explained through a perturbative picture. The standard approach would be to treat $\hat{H}_{c, s}$ using degenerate perturbation theory with respect to $\hat{T}$. However, we can improve on this by considering the operator
\beq
    \hat{T}_{0} = \sum_{x, y\neq x} T_{yx} \delta(V_{x} - V_{y}) \ket{y}\bra{x},
\eeq
where $T_{yx} = \braket{y | \hat{T} | x}$, $V_{x} = \braket{x | \hat{V} | x}$ and $\delta(V_{x} - V_{y})$ is the Dirac-delta function.
We then write $\hat{H}_{c, s} = \hat{H}_{0} + e^{-s} \delta \hat{T}_{0}$, where $\hat{H}_{0} = \hat{V} + e^{-s}\hat{T}_{0}$ and $\delta \hat{T}_{0} = \hat{T} - \hat{T}_{0}$.
While it is unusual to have the perturbing parameter within our choice of $\hat{H}_{0}$, it is important to note that $[\hat{T}_{0}, \hat{V}] = 0$ and thus the eigenstates of $\hat{H}_{0}$ are independent of $s$ (but not their respective eigenvalues).
The operator $\hat{T}_{0}$ is sometimes referred to as the {\em folded} model \cite{Zadnik2021,Zadnik2021b}; in the limit of $s \to \infty$, the dynamics of the system within a sector with fixed $\hat{V}$ is entirely determined by $\hat{T}_{0}$.
The name comes from the fact that the bands in the spectrum [as illustrated in Fig.~\ref{fig: folded}(a)] ``fold'' onto one another.

Applying this analysis to the Fredkin model is simple.
One must first find all matrix elements $T_{xy}$ with $V_{x} = V_{y}$.
This can be done at the level of local transitions.
Under the constraint in \er{H}, the transitions are defined by local four-body configurations, see Fig.~\ref{fig: folded}(b). 
Considering each of these transitions, one can then calculate the local contribution to the potential energy before and after the transition.
Indeed, since the transition can alter the kinetic constraint of a neighbouring pair of particles, it is now necessary to consider local six-body configurations (involving the two sites at either side of the pair which undergoes the transition).
Notice that while there are originally three constraints which depend on the neighbouring sites, there are now twelve possibilities which depend on both the neighbouring and next-neighbouring sites.
One must  determine which possibilities allow for transitions with $V_{x} = V_{y}$.
Table~\ref{table: folded} shows each of these possibilities along with their change in potential energy $\Delta V_{\newmoon\fullmoon\rightarrow\fullmoon\newmoon}$ for the hoppings $\newmoon\fullmoon\rightarrow\fullmoon\newmoon$ (the reverse is given by $-\Delta V_{\newmoon\fullmoon\rightarrow\fullmoon\newmoon}$) and the value of $c$ for which $\Delta V_{\newmoon\fullmoon\rightarrow\fullmoon\newmoon} = 0$.

\begin{figure*}[t]
    \centering
    \includegraphics[width=\linewidth]{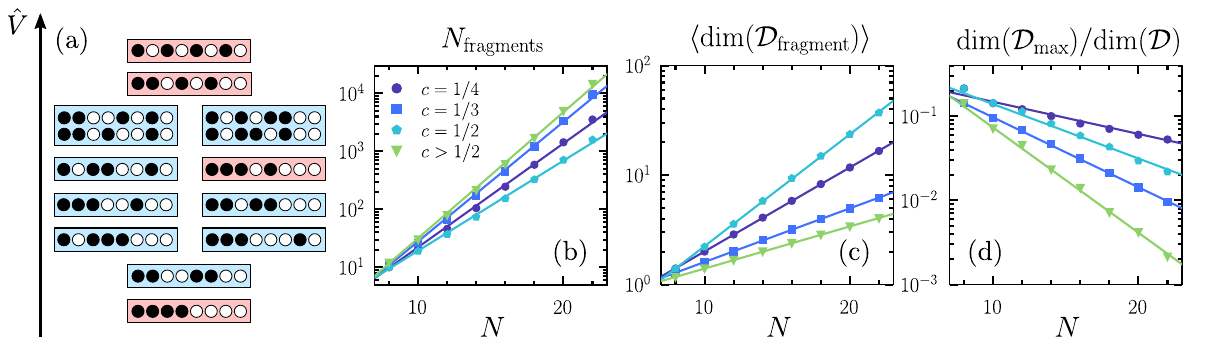}
    \caption{\textbf{Hilbert space fragmentation in the folded model.}
    (a) An illustration of the fragmentation for $N = 8$ and $c > 1/2$.
    Each block describes a sector within $\mathcal{D}$; the red blocks are the isolated configurations $\ket{P_{j}}$.
    The vertical position of each block represents the value of the conserved quantity $\hat{V}$ for each sector.
    (b) The number of fragments within $\mathcal{D}$ grows exponentially with $N$.
    (c) The average dimension of each ergodic component, $\braket{{\rm dim}(\mathcal{D}_{\rm fragment})}$, within $\mathcal{D}$ also grows exponentially.
    (d) On the contrary, the ratio of the dimension of the largest ergodic component ${\rm dim}(\mathcal{D}_{\rm max}) / {\rm dim}(\mathcal{D})$ decreases exponentially with system size.
    }
    \label{fig: fragmentation}
\end{figure*}

\begin{table}[t]
    \begin{centering}
        \begin{NiceTabular}{ccc}[colortbl-like, hvlines]
            \rowcolor{gray!15}
            Transition &  \, $\Delta V_{\newmoon\fullmoon\rightarrow\fullmoon\newmoon}$ \,  & \, Folded \, \\ 
            $\newmoon\newmoon\newmoon\fullmoon\newmoon\newmoon \leftrightarrow \newmoon\newmoon\fullmoon\newmoon\newmoon\newmoon$ & $0$ & All $c$ \\
            $\newmoon\newmoon\newmoon\fullmoon\newmoon\fullmoon \leftrightarrow \newmoon\newmoon\fullmoon\newmoon\newmoon\fullmoon$ & $1 - 2c$ & $c = 1/2$ \\
            $\newmoon\newmoon\newmoon\fullmoon\fullmoon\newmoon \leftrightarrow \newmoon\newmoon\fullmoon\newmoon\fullmoon\newmoon$ & $4c - 1$ & $c = 1/4$ \\
            $\newmoon\newmoon\newmoon\fullmoon\fullmoon\fullmoon \leftrightarrow \newmoon\newmoon\fullmoon\newmoon\fullmoon\fullmoon$ & $-2c$ & None \\
            $\newmoon\fullmoon\newmoon\fullmoon\fullmoon\newmoon \leftrightarrow \newmoon\fullmoon\fullmoon\newmoon\fullmoon\newmoon$ & $0$ & All $c$ \\
            $\newmoon\fullmoon\newmoon\fullmoon\fullmoon\fullmoon \leftrightarrow \newmoon\fullmoon\fullmoon\newmoon\fullmoon\fullmoon$ & $2c-1$ & $c = 1/2$ \\
            $\fullmoon\newmoon\newmoon\fullmoon\newmoon\newmoon \leftrightarrow \fullmoon\newmoon\fullmoon\newmoon\newmoon\newmoon$ & $2c-1$ & $c = 1/2$ \\
            $\fullmoon\newmoon\newmoon\fullmoon\newmoon\fullmoon \leftrightarrow \fullmoon\newmoon\fullmoon\newmoon\newmoon\fullmoon$ & $0$ & All $c$ \\
            $\fullmoon\newmoon\newmoon\fullmoon\fullmoon\newmoon \leftrightarrow \fullmoon\newmoon\fullmoon\newmoon\fullmoon\newmoon$ & $6c-2$ & $c = 1/3$ \\
            $\fullmoon\newmoon\newmoon\fullmoon\fullmoon\fullmoon \leftrightarrow \fullmoon\newmoon\fullmoon\newmoon\fullmoon\fullmoon$ & $4c-1$ & $c = 1/4$ \\
            $\fullmoon\fullmoon\newmoon\fullmoon\fullmoon\newmoon \leftrightarrow \fullmoon\fullmoon\fullmoon\newmoon\fullmoon\newmoon$ & $2c-1$ & $c = 1/2$ \\
            $\fullmoon\fullmoon\newmoon\fullmoon\fullmoon\fullmoon \leftrightarrow \fullmoon\fullmoon\fullmoon\newmoon\fullmoon\fullmoon$ & $0$& All $c$ \\
        \end{NiceTabular}
    \end{centering}
    \caption{\label{table: folded} {\bf Allowed transitions in the folded Fredkin model.} 
    The constraint of the folded model depends on both the nearest and next-nearest neighbours of the sites making the transition: the left column shows all such neighbourhoods, including only those where the constraint of the original model is also satisfied, see Fig.~\ref{fig: folded}(b).
    The central column shows the change in the potential energy due to each transition. The right column shows the values of $c$ for which the transitions are resonant, and therefore possible in the corresponding folded model. 
    }
\end{table}

\subsection{Effective models}
There are four possibilities which are allowed for {\em all} values of $c$, see Tab.~\ref{table: folded}.
Together, these four possibilities can be collectively described by an exclusive-NOR (XNOR) constraint on the neighbouring sites, and an XNOR constraint on the next-nearest neighbouring sites, where the XNOR constraint is one only if the two sites take the same value.
That is, the constraint is activated if the left neighbouring site is in the same state as the right neighbour to the pair, and the 
left next-nearest neighbour is in the same state as the right next-nearest neighbouring site (but note however that the neighbouring sites can differ from the next-nearest neighbouring sites).
This is illustrated in Fig.~\ref{fig: folded}(c).

For the special cases $c = 1/2, 1/3, 1/4$, there are additional allowed moves.
The least constrained case is $c = 1/2$, which allows for a total of eight possibilities which can be summarised by the XNOR constraint {\em only} on the neighbouring sites, shown in Fig.~\ref{fig: folded}(d).
This dynamics (sometimes called the ``folded XXZ'' model) has been considered in various studies, see Refs.~\cite{Yang2020, Pozsgay2021, Singh2021, Zadnik2021, Zadnik2021b, Bastianello2022, Zadnik2023}.
Furthermore, this constraint is intimately related to the generalised Fredkin spin chains studied in Ref.~\cite{Langlett2021}, which considered \er{H} with $s = 0$, but with a kinetic constraint $\hat{n}_{i-1} - \hat{v}_{i+2}$.
The introduction of the minus sign causes the operators to act destructively;
while not identical to the folded Fredkin model for $c=1/2$, it is important to note the similarity between the effective constraints.
The form of the constraint for the special cases $c = 1/3$ and $c = 1/4$ have additional terms which allows for more possibilities for moves, see Figs.~\ref{fig: folded}(e, f). 

The short-time dynamics for the model is then approximately described by the effective Hamiltonian 
\begin{multline}
    \hat{H}^{\rm eff}_{c, s} = -e^{-s}\sqrt{c(1-c)} \sum_{i=2}^{N-2} \hat{g}^{(c)}_{i} \big[ \hat{\sigma}_{i}^{+}\hat{\sigma}_{i+1}^{-}
    + \hat{\sigma}_{i}^{-}\hat{\sigma}_{i+1}^{+}\big]
    \\
    + \sum_{i=2}^{N} \hat{f}_{i}\big[c\hat{v}_{i}\hat{n}_{i+1} + (1-c)\hat{n}_{i}\hat{v}_{i+1}\big],
    \label{H_eff}
\end{multline}
where $\hat{g}^{(c)}_{i}$ are the emergent kinetic constraints shown in Figs.~\ref{fig: folded}(c-f), where for $c \neq 1/2, 1/3, 1/4$,
\begin{align}
    \hat{g}_{i}^{(c)} &= 
    \hat{g}_{i}
    = 
    \frac{1}{4}(1 + \sigma_{i-1}^{z}\sigma_{i+2}^{z})(1 + \sigma_{i-2}^{z}\sigma_{i+3}^{z}),
    \label{g}
\end{align}
while for the special cases $c = 1/2$, $c = 1/3$ and $c = 1/4$,
\begin{align}
    \hat{g}^{(1/2)}_{i} &= \frac{1}{2}(1 + \sigma_{i-1}^{z}\sigma_{i+2}^{z}),
    \\
    \hat{g}^{(1/3)}_{i} &= \hat{g}_{i}  + 2\hat{v}_{i-2}\hat{n}_{i-1}\hat{v}_{i+2}\hat{n}_{i+3},
    \\
    \hat{g}^{(1/4)}_{i} &= \hat{g}_{i}
    + 2(\hat{n}_{i-2}\hat{n}_{i-1}\hat{v}_{i+2}\hat{n}_{i+3} 
    \\
    & \;\;\;\;
    + \hat{v}_{i-2}\hat{n}_{i-1}\hat{v}_{i+2}\hat{v}_{i+3}).
    \nonumber
\end{align}
The main focus of what follows will be the case $c > 1/2$, which results in the most constrained dynamics with the constraint \er{g}, see Figs.~\ref{fig: folded}(c). 
As we show below, the existence of the folded model in the $s \to \infty$ limit has dramatic consequences for the dynamics at finite $s > 0$.

\subsection{Fragmentation}

The effective Hamiltonian $\hat{H}^{\rm eff}_{c, s}$ is a more constrained model than \er{H} which describes its leading order dynamics. The more stringent kinetic constraint of the folded model has severe consequences on its dynamics (and thus on the short time dynamics of the full model). One important feature of \er{H_eff} is that the Hilbert space is fragmented: the subspaces of the original Hamiltonian divide into smaller subspaces \cite{Moudgalya2022, Moudgalya2022b}.

For concreteness, let us focus on the largest subspace of the Fredkin chain, $\mathcal{D}$.
Figure~\ref{fig: fragmentation}(a) illustrates how  it fragments for a system size $N = 8$ and $c > 1/2$, for which $\dim\mathcal{D}=14$.
Each block in the figure is an ergodic component. Notice that the subspace strongly fractures into many components, with the majority being an isolated computational basis state, and with at most two states for the system size shown here.
In particular, for $c \neq 1/4, 1/3, 1/2$, the product states $\ket{P_{j}}$ are completely isolated, and are shown by the red blocks in Fig.~\ref{fig: fragmentation}(a).
Indeed, we can quantify the strength of the fragmentation by counting the number of ergodic components of the folded model within $\mathcal{D}$. This is shown in Fig.~\ref{fig: fragmentation}(b) as a function of $N$ for various 
values of $c$ in Fig.~\ref{fig: folded}. In all cases, the number of ergodic components of the corresponding folded model grows exponentially in $N$.

It is also useful to consider how the size of each ergodic component grows.
It would be reasonable to conclude from Figs.~\ref{fig: fragmentation}(a, b) that $\mathcal{D}$ fragments into sectors which are not extensive in system size.
However, as shown in Fig.~\ref{fig: fragmentation}(c), the mean dimension of the fragments, $\braket{{\rm dim}(\mathcal{D}_{\rm fragment})}$, also grows exponentially but at a rate smaller than that of $\mathcal{D}$.
Figure~\ref{fig: fragmentation}(d) shows the ratio ${\rm dim}(\mathcal{D}_{\rm max}) / {\rm dim}(\mathcal{D})$ where $\mathcal{D}_{\rm max}$ is the dimension of the largest fragment.
Note that this is exponentially decaying to zero with $N$, and thus the space is strongly fragmented \cite{Moudgalya2022}.

While the effective Hamiltonian $\hat{H}_{c, s}^{\rm eff}$ only gives the leading order of $\hat{H}_{c, s}$, the full dynamics can be retrieved perturbatively.
It is important to note that while the eigenstates of $\hat{H}_{c, s}^{\rm eff}$ are by no means trivial, the fragmented Hilbert space allows one to diagonalise individual fragmented sectors for much larger system sizes.
We can then approximately recover the eigenstates of $\hat{H}_{c, s}$ by means of perturbation theory, which we will apply in Sec.~\ref{sec: spectrum}; see App.~\ref{appendix: perturbation} for details.

\section{Ergodicity breaking} \label{sec: relaxation}
In this section, by means of ED, 
we provide numerical evidence for the dynamics of \er{H} being non-ergodic for special initial states, and explain the connection of this phenomenon to the folded model.

\subsection{Level statistics}

We first investigate the level spacing statistics of the Hamiltonian $\hat{H}_{c, s}$ in search of indications of quantum chaos (i.e., Wigner-Dyson level statistics) or  integrability (i.e., Poisson level statistics); for a review, see e.g.\ Ref.~\cite{DAlessio2016}.
For the eigenenergies $E_{n}$ of $\hat{H}_{c, s}$ we calculate the level spacings, $\delta_{n} = E_{n+1} - E_{n}$, normalised such that $\braket{\delta} = 1$.
We obtain $\{E_{n}\}$ using ED for system sizes up to $N = 22$ within the subspace $\mathcal{D}$ and the ${\rm CP} = +1$ sector (with a subspace dimension of $29264$), restricting to the middle $2/3$ of the energy spectrum.

\begin{figure}[t]
    \centering
    \includegraphics[width=\linewidth]{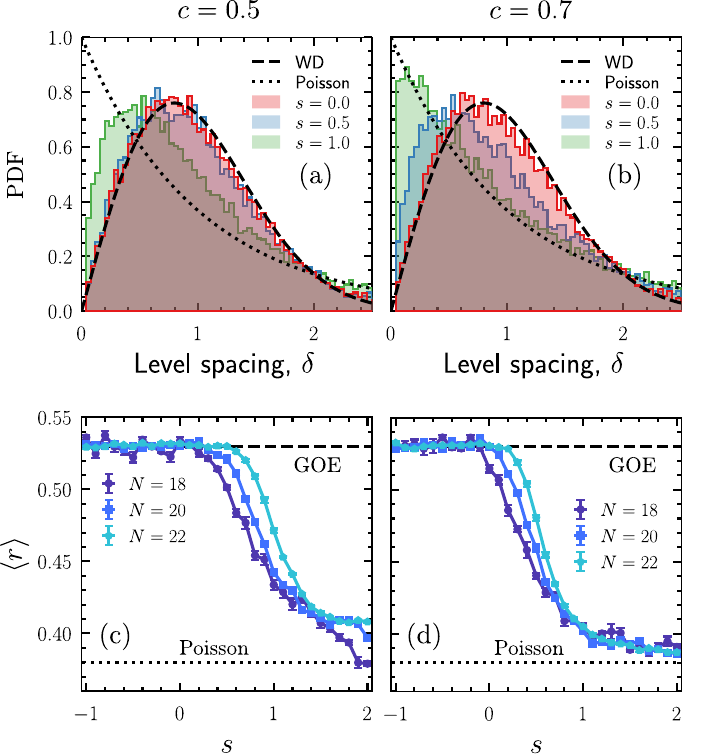}
    \caption{\textbf{Level spacing statistics.} 
    (a) Histograms of the normalised level spacings, $\delta$, at $c=1/2$ in the CP sector ${\rm CP} = +1$ and for system size $N = 22$ (i.e., over an irreducible space with dimension $29624$). We show results for 
    $s = 0$ (red), $s=0.5$ (blue) and $s=1$ (green).
    The dashed line is the Wigner-Dyson (WD) distribution for level repulsion, while the dotted line is for Poisson statistics.
    (b) The same for $c = 0.7$. 
    (c) The average $r$-value, $\braket{r}$, for $c=1/2$
    as a function of $s$ in the sector ${\rm CP} = +1$ for sizes $N = 18, 20, 22$. The dashed line corresponds to level repulsion obtained from the Gaussian orthogonal ensemble (GOE), $\braket{r} \approx 0.53$, and the dotted line to Poisson statistics, $\braket{r} \approx 0.38$.
    Each data point shows the mean $r$-value averaged over five points in the range $[s-\Delta s, s+\Delta s]$ for the indicated $s$ value and the $\Delta s = 0.02$.
    The errors bars show the standard error for the mean.
    (d) The same for $c = 0.7$.
    }
    \label{fig: level_statistics}
\end{figure}

Figures \ref{fig: level_statistics}(a, b) show the level spacing distributions for $c=1/2$ and $c = 0.7$, respectively. Each panel shows the statistics for $s = 0, 0.5, 1.0$, comparing to the Wigner-Dyson (WD) distribution (dashed line), and to the Poisson distribution (dotted line). We find that for $s \lesssim 0$, the statistics are consistent with Wigner-Dyson, indicating the usual level  repulsion. In contrast, for increasing $s \gtrsim 0$ there appears to be a shift towards Poisson statistics, which is more pronounced for $c>1/2$.

The above observations can be condensed by calculating the ratio between consecutive gaps $r$ \cite{Oganesyan2007}:
for each eigenstate, we calculate the quantity $r_{n} = \min\{\delta_{n+1}, \delta_{n}\} / \max\{\delta_{n+1}, \delta_{n}\}$, and then average over all $r_{n}$ to obtain $\braket{r}$
\footnote{For the small system sizes accessible, it can be preferable to truncate the energy spectrum to a central portion, as the small density of states at the edge of the spectrum can result in strong outliers in $r_{n}$. Here, we use only the eigenstates from the centermost $2/3$ of the energy spectrum $E_{n}$.}.
This takes the value $\braket{r} \approx 0.53$ for the Gaussian orthogonal ensemble (Wigner-Dyson distribution), and $\braket{r} \approx 0.38$ for the uncorrelated spectrum (Poisson distribution).
In Figs.~\ref{fig: level_statistics}(c, d) we show $\braket{r}$ for the same systems above for sizes $N = 18, 20, 22$. 

In the limit $s\to\infty$, the Hamiltonian is well described by \er{H_eff} and there are additional (emergent) kinetic constraints, which explains the apparent Poisson statistics of the level spacing (dotted line) for $s>0$. In turn, for $s \lesssim 0$, the observed value is more compatible with the Wigner-Dyson statistics (dashed line) one would expect to see in ergodic systems. We note that for the case of $c=1/2$, the change from Wigner-Dyson to Poisson statistics seems to be lessened with increasing system size, with the change occuring at an $s$ that increases with $N$.
A similar observation can be made for $c = 0.7$.
In this case, the crossover looks to be getting sharper with increasing system size.
However, it is hard to conclude the nature of this change from the small system sizes that are accessible.
Nevertheless, the key observation is that there is a significant change in statistics close to the stochastic point $s = 0$.

\subsection{Prethermalisation}

Next we show that the change described above in the spectrum is related to slow relaxation. In order to probe metastability, we consider the time-averaged autocorrelations of the site occupations $\hat{n}_{i}$, with respect to the infinite temperature state within the sector $\mathcal{D}$.
We define the autocorrelation of the lattice site $i$ at some time $t$ as $c_{i}(t) = \braket{\hat{n}_{i}(t)\hat{n}_{i}(0)}_{\mathcal{D}}$, where the subscript denotes the expectation is taken with respect to the infinite temperature state within $\mathcal{D}$.
By summing over lattice sites
and taking the time-average, we then find
\beq
    \overline{c(t)} = t^{-1}\sum_{j=1}^{N}\int_{0}^{t} dt' c_{j}(t').
\eeq
Finally, we normalise,
\beq
    \overline{C(t)} = \frac{\overline{c(t)} - \overline{c(\infty)}}{\overline{c(0)} - \overline{c(\infty)}},
\eeq
such that $C(0) = 1$ and $C(\infty) = 0$.

\begin{figure}[t]
    \centering
    \includegraphics[width=\linewidth]{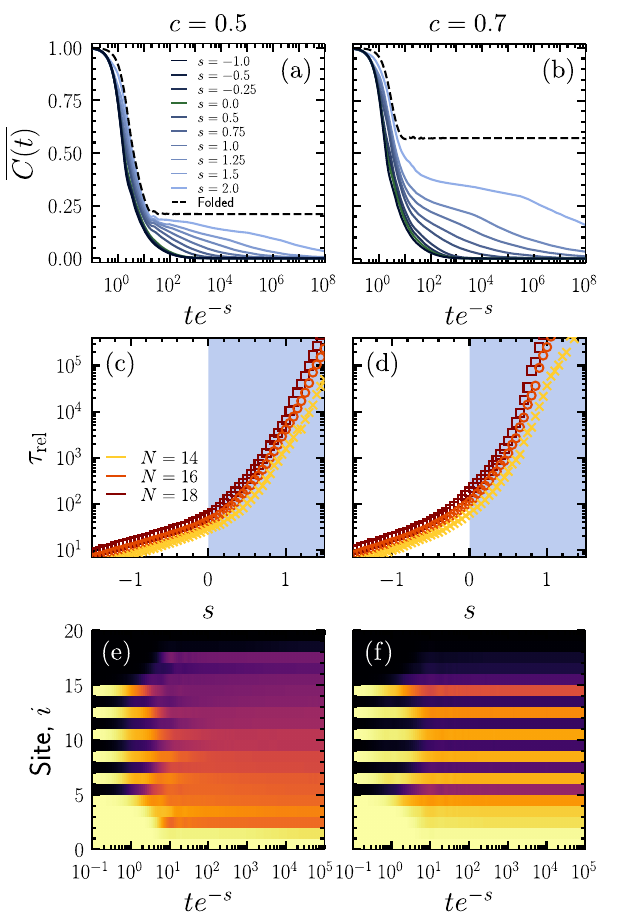}
    \caption{\textbf{Relaxation dynamics.} 
    (a) The normalised (and time-integrated) autocorrelation function $\overline{C(t)}$ of the site occupations at $c=1/2$ for various $s$ and $N = 18$. 
    The dashed black lines show the correlation function under the folded dynamics \er{H_eff} at $s = 2.0$.
    (b) The same for $c= 0.7$, where the plateaus are more pronounced. 
    (c) The relaxation timescale $\tau_{\rm rel}$ as a function of $s$ for system sizes $N = 14, 16, 18$ and $c = 0.5$.
    (d) The same for $c = 0.7$.
    (e) Time-averaged occupation profiles for $s = 1.0$ and $N = 20$ starting from the initial state $\ket{P_{N/4}}$. 
    (f) The same for $c= 0.7$. 
    Note that time is rescaled by $e^{-s}$ as this is the bare timescale for the kinetic energy, cf.~\eqref{H}.
    }
    \label{fig: autocorrelations}
\end{figure}

Figures~\ref{fig: autocorrelations}(a, b) show the autocorrelation functions for $c=1/2$ and $c = 0.7$ at system size $N = 18$ and for various $s$.
In both instances, for $s\gtrsim 0$, there is an initial decay, followed by a plateau with a much longer life-time, which becomes more pronounced with increasing $s$.
This behaviour is easily explained using the folded picture.
At early times, the dominant contributions to the dynamics comes from the effective Hamiltonian, \er{H_eff}.
This is shown for $s = 2$ by the dashed black line.
Note that we normalise it with respect to the full Hamiltonian, \er{H}, and thus it does not decrease to zero in the infinite time limit.
The two-point correlator measured from the effective Hamiltonian well matches the true dynamics at short times.

For $c=1/2$, the folded dynamics stops being a reliable description at the point when the autocorrelation plateaus.
It is at this point that the system has thermalised within a sector of the folded model, and the long-time dynamics is dominated by the off-resonance transitions not described by it. The relaxation for $c = 0.7$ is much slower and becomes more pronounced with increasing $s$ due to the strong fragmentation in the folded model, cf.\ Figs.~\ref{fig: folded}(c,d). 

From the autocorrelation functions, we are able to measure a relaxation timescale $\tau_{\rm rel}$, which we estimate as the first time when $\overline{C(t = \tau_{\rm rel})}$ falls below $0.05$.
We show this as a function of $s$ in Figs~\ref{fig: autocorrelations}(c, d) for system sizes $N = 14, 16, 18$.
Notice that for $c = 0.5$, it is clear that there is a change in behaviour around the stochastic point $s = 0$, with the timescale increasing far more rapidly for $s > 0$.
For $c = 0.7$, this change is less clear: there appear to be multiple kinks in the curves, for both $s < 0$ and $s > 0$, indicating a competition of multiple timescales. 
Nevertheless, it is clear that there is still a change from a fast-thermalizing regime to a slow-thermalizing regime around the stochastic point.

The long-relaxation times can also be seen by considering the evolution of local occupations,
\beq
    \overline{\braket{\hat{n}_{j}(t)}} = t^{-1}\int_{0}^{t} dt' \braket{\psi(t') |\hat{n}_{j} | \psi(t')}
\eeq
with respect to some initial state $\ket{\psi(0)}$, 
where we have once again taken the time-average to remove the effects of short scale fluctuations.
We show the occupation profiles for the initial state $\ket{P_{N/4}}$ in Figs.~\ref{fig: autocorrelations}(e, f) for $s = 1.0$ and $N = 20$.
For $c=1/2$, the system remembers its initial conditions for intermediate times, eventually relaxing to a state in which the initial density modulations are removed, 
see Figs.~\ref{fig: autocorrelations}(e). 
In contrast, for $c = 0.7$ the initial density pattern persists for the longest simulated times, 
see Figs.~\ref{fig: autocorrelations}(f). 
This difference is easy to understand from the fact that 
for $c = 0.7$ the initial state is a frozen irreducible configuration in the folded picture. This is not the case for $c=1/2$, as for this value of $c$ there are allowed  on-resonance processes in the folded dynamics that lead to relaxation to an almost featureless state within the simulated timescales.

\subsection{Entanglement entropy dynamics}

\begin{figure*}[t]
    \centering
    \includegraphics[width=\linewidth]{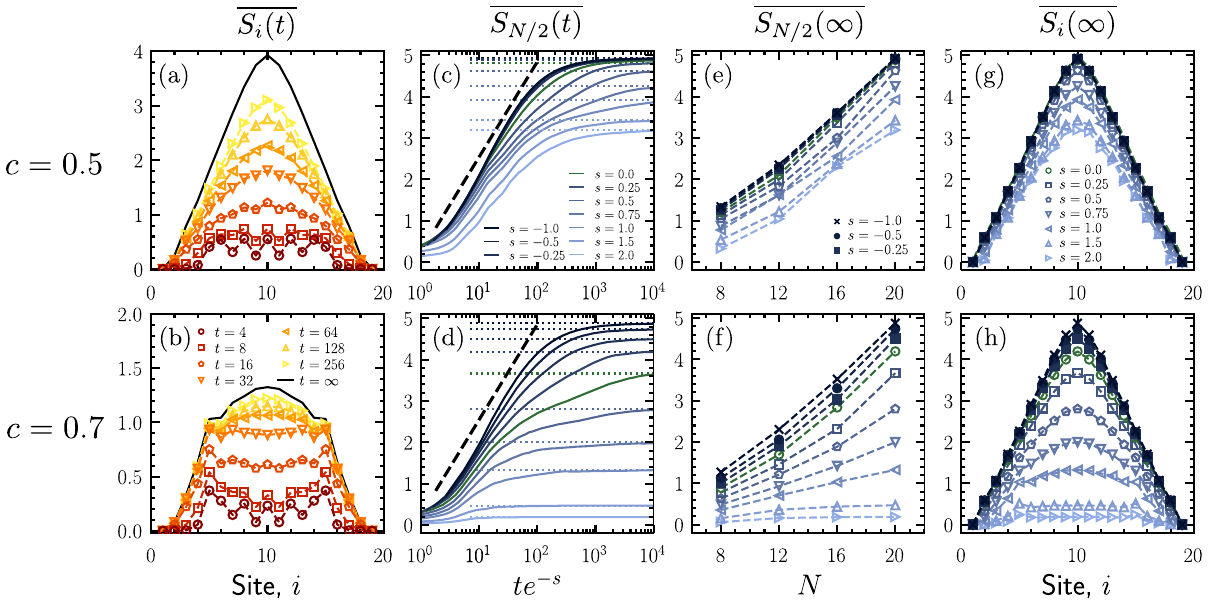}
    \caption{
        \textbf{Entanglement dynamics.} 
        Dynamics following a quench from the state $\ket{P_{N/4}}$. (a) The time-averaged entropy profiles $\overline{S_{i}(t)}$ at $c=1/2$ for $s = 1.0$ and $N = 20$, at various times $t$. The solid black line is the estimated value for $t = \infty$ (obtained by time-averaging in the range $t \in [1 \times 10^{12}, 2 \times 10^{12}]$). (b) The same for $c=0.7$. (c) The time-averaged entanglement entropy for the half system bipartition, $\overline{S_{N/2}(t)}$, as a function of time (scaled by $e^{-s}$) for various $s \in [-1.0, 2.0]$. The dashed black line shows logarithmic growth, and the dotted lines give the saturation values at $t\to\infty$. (d) The same for $c=0.7$. (e) Long-time averaged $\overline{S_{N/2}(\infty)}$ as a function of system size for the same values of $s$. (f) The same for $c=0.7$ 
        (the apparent super-linear in $N$ growth of $s \lesssim 0.5$ is likely to be an artefact of the small systems sizes studied). (g) Long-time averaged entanglement profiles $\overline{S_{i}(\infty)}$ for the same values of $s$. (h) The same for $c=0.7$.
    }
    \label{fig: dynamics}
\end{figure*}

The effect of the constrained dynamics can be further understood by considering the growth of entanglement entropy for simple initial states. Specifically, given a state $\ket{\psi(t)}$ at time $t$, we partition the system into parts $A$ and $B$ with the cut taken between sites $i$ and $i+1$. We then calculate the bipartite entanglement entropy between $A$ and $B$, 
\beq
    S_{i}(t) = -\Tr_{A} \left[\rho_{A}(t) \ln\rho_{A}(t)\right],
\eeq
where $\rho_{A}(t) = \Tr_{B} \ket{\psi(t)}\bra{\psi(t)}$.
To consider the time evolution of the entanglement entropy, as before we perform the time average to smooth out uninteresting fluctuations,
\beq
    \overline{S_{i}(t)} = t^{-1}\int_{0}^{t} dt' \, S_{i}(t').
\eeq
Figure~\ref{fig: dynamics} shows such time-integrated  entanglement entropy for the product state $\ket{P_{N/4}}$, see \er{Pj}, at $c=1/2$ in the top row, and at $c=0.7$ in the bottom row.

We first consider the evolution of the entanglement entropy profiles for all bipartitions $i$ at different times, see Figs.~\ref{fig: dynamics}(a,b). The entanglement profiles are heterogeneous for small times in both cases, where the dynamics is approximately described by evolution with the folded \er{H_eff}. For $c=1/2$, the constraint of the folded model is that of Fig.~\ref{fig: folded}(d). For states of the form $\ket{P_{j}}$ this implies that all sites are frozen except at the two bonds at the interfaces of the three domains that define $\ket{P_{j}}$, cf.\ \er{Pj}. The initial dynamics generated here is then able to spread allowing the state to relax within the sector of the model. For $c>1/2$ the situation is slightly different.
For the folded model, Fig.~\ref{fig: folded}(c), the state $\ket{P_{j}}$ is a dynamically frozen configuration. A transition from the full model is required near the boundaries of the domain to get the dynamics going.
Once this occurs, the folded constraint Fig.~\ref{fig: folded}(c) can be satisfied at the boundary to allow on resonance transitions there, see Tab.~\ref{table: folded}. From Fig.~\ref{fig: dynamics}(b), it is clear that at early times the entanglement grows quickest at these boundaries.

Figures~\ref{fig: dynamics}(c, d) show the growth of entanglement with time at for the midpoint bipartition, $\overline{S_{N/2}(t)}$. 
The dotted lines are the estimated long-time averaged entropy.
In both cases, there is a slow logarithmic growth of entanglement, before eventually saturating to some long-time limit. For $c=1/2$, the effect of increasing $s$ is small, as the folded dynamics allows $\ket{P_{j}}$ to thermalise within a fragmented sector that is extensive in $N$.
The same is not true for $c = 0.7$, where the initial state is frozen, and the saturation value decreases sharply with $s$.
Note that while this decrease in saturation values occurs for all $s$ shown, it is most profound for $s > 0$.

Figure~\ref{fig: dynamics}(e) shows that for $c=1/2$
the saturation value of the entropy grows as a function of $N$ for all $s$.
This fact suggests the lack of a sharp change in behaviour for $c=1/2$ when going from $s$ positive to $s$ negative (despite the fact that there is a small decrease in the saturation value of the entropy at fixed $N$ for increasing $s$).
In contrast, Fig.~\ref{fig: dynamics}(f) shows that at $c = 0.7$ the entropy saturates with $N$ for large enough $s>0$.
Again, this is explained through the folded model. [Note that for small $s > 0$ and $c = 0.7$, the entanglement appears to grow with system size, but we expect it will eventually saturate for larger $N$.]

For $s \lesssim 0$, we observe a linear growth of the entropy with $N$, with similar slope for all the curves, suggesting a volume law, even though a definite conclusion cannot be drawn from the limited system sizes available. 
Figures~\ref{fig: dynamics}(g, h) show the long-time averaged entanglement entropy profiles for all bipartitions. For $c=1/2$, they take a typical structure, with the entropy increasing approximately linearly up to the midpoint of the system. For $c = 0.7$, the entropy profiles shows a larger spatial variations, which become more pronounced for increasing $s$.

\section{Nonthermal eigenstates} \label{sec: spectrum}
The folded model and the observed slow heterogeneous dynamics suggests the existence of non-thermal behaviour at the level of the spectrum. In this Section we investigate the spectral properties of the model by means of ED, perturbation theory and variational MPS.
We verify the existence of the non-thermal eigenstates far from the large interaction limit, and for large system sizes.

\subsection{Spectral properties from exact diagonalisation}

\begin{figure*}[t]
    \centering
    \includegraphics[width=\linewidth]{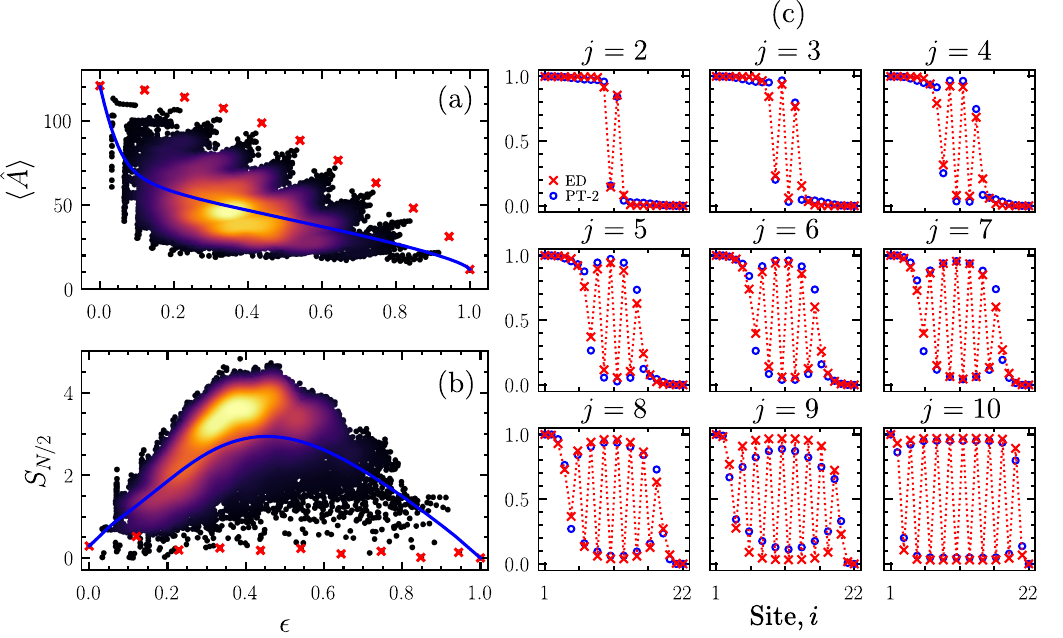}
    \caption{\textbf{Spectrum and non-thermal eigenstates.}
    (a) Expectation of the area, $\braket{A}$, as  a function of the renormalised energy density, $\epsilon$, 
    for $c = 0.7$ and $s = 0.8$ at size $N = 22$ and ${\rm CP} = +1$ from ED. The colour indicates the density of states in the local neighbourhood (with lighter colour corresponding to higher density). Red crosses are the non-thermal eigenstates $\ket{{\mathcal S}_{j}}$ for $j = 1, \dots, N/2$ (left-to-right). 
    The blue line is the canonical average.
    (b) Same for the entanglement entropy for the  mid-point bipartition.
    (c) Density profiles in the non-thermal eigenstates (red crosses),
    $\braket{{\mathcal S}_{j} | n_{i} | {\mathcal S}_{j}}$, for $j=2, \dots, N/2 - 1$. We also show the same profiles in the approximate eigenstate from second-order perturbation theory (blue circles). 
    \label{fig: spectrum}}
\end{figure*}

We use ED to determine all the energy eigenstates in the subspace $\mathcal{D}$.
By exploiting the symmetries of the model, we are able to run the calculations for system sizes up to $N = 22$.
Figure~\ref{fig: spectrum} shows the results of ED for $c = 0.7$, $s = 0.8$, $N = 22$ in the sector ${\rm CP} = +1$.
As a relevant observable we consider the expectation of the area, defined in Eq.~\eqref{eq:area}, within each eigenstate $\braket{\hat{A}}$,  see Fig.~\ref{fig: spectrum}(a). The thermal (canonical) average within the subspace $\mathcal{D}$,
\beq
    \braket{\hat{A}}_{\rm \beta} = \frac{\sum_{E_{j}} e^{-\beta E_{j}} \braket{E_{j} | \hat{A} | E_{j}}} {\sum_{E_{j}} e^{-\beta E_{j}}},
\eeq
at the inverse temperature $\beta$ that corresponds to the same energy density,
is also shown by the solid blue line for comparison.
It is apparent that there are a number of eigenstates with an expectation of the area far from the thermal value, an indication that ETH might be violated by these states.
The largest discrepancy is observed by the states ($N/2$ in total) marked by the red crosses, which appear to be equispaced in energy. In analogy to recent results \cite{Turner2018}, we interpret these as {\em scarred eigenstates}, $\ket{{\mathcal S}_{j}}$, with $j = 1, \cdots, N/2$
\footnote{
    The states $\ket{{\mathcal S}_{1}}$ and $\ket{{\mathcal S}_{N/2}}$ are not strictly non-thermal eigenstates, as they correspond to the extreme eigenstates of the Hamiltonian. Since they are at the edge of the spectrum, they also show ``thermal'' values for local observables. We nevertheless include them in the sequence of eigenstates as they fit into the same spatial pattern.}.

The (midpoint) bipartite entanglement entropy, $S_{N/2}$, of all eigenstates is shown in Fig.~\ref{fig: spectrum}(b).
In analogy with the area, a significant number of eigenstates, including the non-thermal eigenstates $\ket{{\mathcal S}_{j}}$, have a value of this entropy which is smaller than the one obtained by averaging over a small energy window. For the purpose of comparison, we can define as a proxy its ``thermal'' average as
\beq
    S_{\rm \beta} = \frac{\sum_{E_{j}} e^{-\beta E_{j}} S_{N/2}(E_{j})} {\sum_{E_{j}} e^{-\beta E_{j}}},
\eeq
where $S_{N/2}(E_{j})$ is the bipartite entanglement entropy of the individual eigenstate, $\ket{E_{j}}$. Figure~\ref{fig: spectrum}(b) shows that this proxy (blue curve) differs from the actual entropies of the non-thermal eigenstates (red crosses).

To further understand the properties of the non-thermal eigenstates, we calculate their local observables.
In particular, we measure the local occupation profiles $\braket{n_{i}} = \braket{{\mathcal S}_{j} | n_{i} | {\mathcal S}_{j}}$ for lattice sites $i = 1, \dots, N$, shown by the red crosses in Fig.~\ref{fig: spectrum}(c) for $j = 2, \dots, N/2-1$.
Notice that the structure of each eigenstate can be separated into approximately three partitions: the first is $N/2+1-j$ spins that have a large occupation for spin-up, the second has $2j-2$ spins that are approximately antiferromagnetic, and the third is $N/2+1-j$ spins that have a large occupation for spin-down.
These highly resemble the product states \er{Pj}, indicating each $\ket{\mathcal{S}_{j}}$ have a high overlap with $\ket{P_{j}}$.
We verify this connection using second order perturbation theory.
In the limit $s\to\infty$, each of the product states $\ket{P_{j}}$ are eigenstates of the Hamiltonian \er{H}.
The eigenstates for $s\neq\infty$ can be estimated using perturbation theory on the states $\ket{P_{j}}$, see App.~\ref{appendix: perturbation} for more details, and are shown by the blue circles in Fig.~\ref{fig: spectrum}(c).
Note that for the most part, the results of perturbation theory closely agree with those of ED.
Even for the case of $j=9$ where the disagreement is the largest, the results qualitatively match, indicating that perturbation theory can help explain the dynamics of the model, even when far from the $s=\infty$ limit.

\subsection{Extracting non-thermal eigenstates with MPS}

We now seek to investigate the non-thermal eigenstates for system sizes larger than those accessible via ED. In particular, the low-entanglement properties of the non-thermal eigenstates states suggest that an MPS approximation might be a suitable ansatz for the wavefunctions. While it could be that MPS are able to describe such states, variationally targeting these excited states using methods such as the density matrix renormalisation group (DMRG) \cite{White1992} --- which are best suited for extremal eigenstates --- is difficult due to the fact they exist throughout the entire spectrum. 
To address this issue we use the approach similar to that described in Ref.~\cite{Banuls2020}, which aims to variationally minimise the energy variance of a wavefunction $\psi$,
\beq
    \delta {E_{\psi}}^{2} = \frac{\braket{\psi | \hat{H}^{2} | \psi}}{\braket{\psi | \psi}} - \frac{\braket{\psi | \hat{H} | \psi}^2}{\braket{\psi | \psi}^2},
    \label{cost}
\eeq
using gradient decent to optimise the tensors in a DMRG-like fashion
\footnote{
    An alternative approach for estimating non-thermal eigenstates with MPS worth mentioning is Ref.~\cite{Zhang2023}, which uses a shift-invert like method to estimate eigenstates at a particular energy.
}. 
Reference~\cite{Banuls2020} minimised \er{cost} with the addition of a Lagrange multiplier to target eigenstates at some desired energy. Here, we adapt this variational minimisation to target states which have a large overlap with $\ket{P_{j}}$, see App.~\ref{appendix: vmps} for more details.
Using this algorithm, we are able to estimate to good precision the non-thermal eigenstates for system sizes up to $N = 100$ for $c \geq 0.7$ and $s \gtrsim 0.75$, where the eigenstates are distinguished enough from the bulk spectrum to allow us to target them with this approach.

\subsection{Properties of the non-thermal eigenstates}

\begin{figure*}[t]
    \centering
    \includegraphics[width=\linewidth]{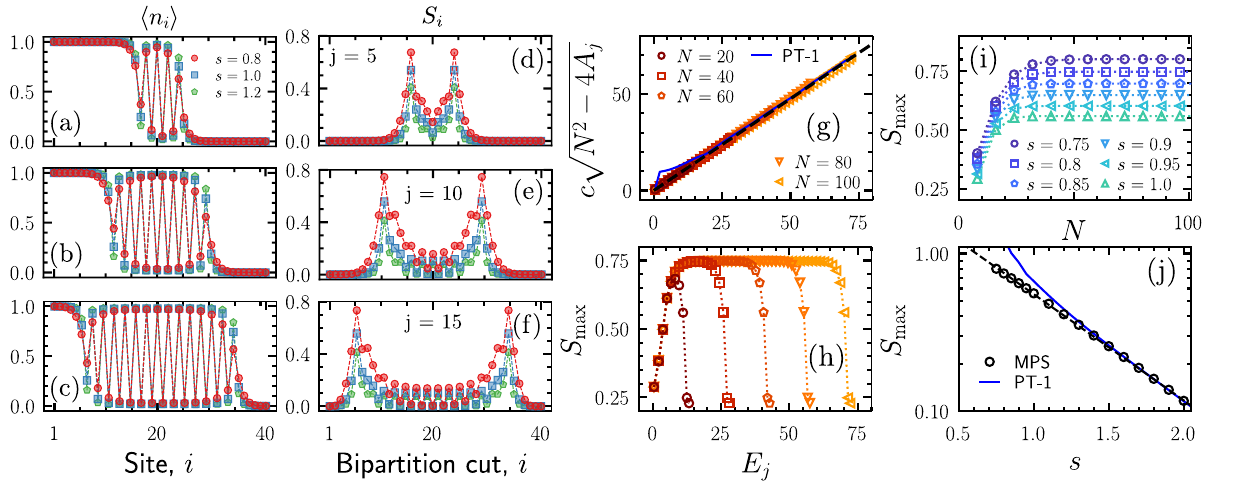}
    \caption{\textbf{Properties of the scarred states found with vMPS.}
    (a-c) Density profiles 
    $\braket{{\mathcal S}_{j}^{\rm MPS} | \hat{n}_{i} | {\mathcal S}_{j}^{\rm MPS}}$, for $j=5,10,15$ (as indicated) for three values of $s$  
    at $c = 0.7$ and system size $N = 40$.
    (d-f) Corresponding bipartite entanglement entropy profiles, $S_{i}$.
    (g) Square root of the rescaled area, $\tilde{A}_{j} = c^{2}\left(N^{2} - 4A_{j}\right)$, as a function of the scarred eigenstate energy, $E_{j}$, at $c = 0.7$ and $s = 0.8$ for system sizes $N = 20, \dots, 100$.
    The dashed black line shows a linear fit, and the solid blue line shows the results from first-order perturbation theory (PT-1) with $N = 100$.
    (h) Maximum bipartite entanglement entropy for the scarred states, $S_{\rm max} = \max_{i} S_{i}$, shown for each $\ket{{\mathcal S}_{j}}$ with $j = 1, \dots, N/2$ for sizes $N = 20, \dots, 100$.
    The data in both (g) and (h) are for $c = 0.7$ and $s = 0.8$.
    (i) Maximum entanglement entropy, $S_{\rm max}$, for the non-thermal eigenstate $\ket{{\mathcal S}_{N/4}}$ as a function of system size at $c=0.7$
    for various $s \in [0.75, 1.0]$. 
    (j) Saturation value of the entanglement entropy
    at $c=0.7$  as a function of $s$, obtained from the non-thermal eigenstate $\ket{{\mathcal S}_{N/4}}$ at size $N = 200$. The dashed line show the apparent exponential decay. The blue line is the results from first-order perturbation theory (PT-1) at size $N = 100$.
    }
    \label{fig: mps_scars}
\end{figure*}

Figure~\ref{fig: mps_scars} shows the results from our MPS numerics for $c = 0.7$. As a proof-of-principle, we show the occupation profiles of the non-thermal eigenstates we obtain, $\ket{{\mathcal S}_j^{\rm MPS}}$, in Figs.~\ref{fig: mps_scars}(a-c) for $N = 40$, $s =0.8, 1.0, 1.2$ and $j = 5$ (top panel), $j = 10$ (middle panel) and $j = 15$ (bottom panel).
Their spatial structure resembles that of the exact ${\mathcal S}_j$ for small systems from ED of Fig.~\ref{fig: spectrum}.
Furthermore, it is evident that larger $s$ results in profiles increasingly resemble those of the product states $\ket{P_{j}}$. Figures~\ref{fig: mps_scars}(d-f) show the corresponding entanglement entropy profiles, $S_{i}$. 
The states $\ket{{\mathcal S}_j^{\rm MPS}}$ can be approximately separated into three regions, similar to those of $\ket{P_{j}}$, see \er{Pj}. The first $i = 1, \cdots, N/2 - j + 1$ spins have an occupation profile $\braket{\hat{n}_{i}} \approx 1$, while the last $i = N/2 + j - 1, \cdots, N$ have $\braket{\hat{v}_{i}} \approx 1$.
Both of these regions have approximately zero entanglement with the rest of the system.
In contrast, in the central region, 
$N/2 - j + 2\leq i \leq N/2 + j$,
the spin profile is approximately an antiferromagnetic pattern, with an entanglement profile that alternates between close to zero and a small but non-zero value. This suggests that there is a dimerised structure within this partition, where neighbouring pairs of spins are coupled but do not interact with the other pairs of spins. As Figs.~\ref{fig: mps_scars}(d-f) show, the entanglement entropy is maximal at the interface of two regions of $\ket{{\mathcal S}_j^{\rm MPS}}$.
The explanation for this is the same as for the growth of entanglement entropy for the product states $\ket{P_{j}}$: the off-resonant transitions at the boundaries of the regions give the smallest change in potential energy, and are thus more prominent. 

Next we consider the properties of the non-thermal eigenstates $\ket{{\mathcal S}_j^{\rm MPS}}$ as one increases system size. 
It is easy to show that the product states $\ket{P_j}$ have energy $E_j=\braket{P_{j} | \hat{H} | P_{j}} = 2(1 - cj)$ and area 
\begin{multline}
    \braket{P_{j} | \hat{A} | P_{j}} = N^{2} / 4 - j(j-1)
    \\ 
    =\frac{N^2}{4}-\frac{4(E_j-2+c)^2}{c^2}+4.
\end{multline}
Since the non-thermal eigenstates are very close to these products, we can expect that a similar relation holds between their area and energy. To probe it,
Figure~\ref{fig: mps_scars}(g) shows the square root of the rescaled area for the non-thermal eigenstates, 
$\tilde{A}_{j} = c^{2}(N^{2}-4A_{j})$, 
as a function of energy for $s = 0.75$ and system sizes $N = 20, \dots, 100$ for all non-thermal eigenstates $j = 1, \cdots, N/2$.
It is clear that the quadratic relation $A_{j} \sim E_{j}^{2}$ holds for the non-thermal eigenstates. The figure shows that there is excellent agreement with the result of first-order perturbation theory around $\ket{P_j}$.

We also consider the entanglement entropy as a function of system size. Figure~\ref{fig: mps_scars}(h) shows the maximum bipartite entanglement entropy over all lattice sites, $S_{\rm max} = \max_{i} S_{i}$, for the same non-thermal eigenstates consider in Fig.~\ref{fig: mps_scars}(g). When close to the edges of the spectrum, the maximal entanglement entropy is small, as is expected for local quantum many-body systems. Under the ETH, the entanglement entropy of eigenstates should grow as the energy moves closer to the middle of the spectrum.
However, for the non-thermal eigenstates the maximal entanglement  saturates to a value which appears to be independent of system size (in the large $N$ limit). This suggests that the non-thermal eigenstates obey an area law, thus violating the ETH. 
We confirm this more directly in Fig.~\ref{fig: mps_scars}(i), where we show the maximal entanglement entropy as a function of system size, $N$, for several $s$ for the non-thermal eigenstate $j = N/4$ (the one closest to the middle of the energy spectrum): it is clear that the maximal entanglement entropy saturates with increasing system size for this state. 
Figure~\ref{fig: mps_scars}(j) shows the maximal entanglement entropy of the state $j = N/4$ as a function of $s$ for $N = 200$.
The maximal entanglement entropy appears to decrease exponentially with $s$. 
We also compare this result to first-order perturbation theory for $N = 100$, which coincides with the MPS numerics for large $s$.

\section{Conclusions} \label{sec: conclusions}
Here we studied the quantum dynamics of Fredkin spin chains deformed away from their stochastic point.
The introduction of a ``tilting'' parameter allows us to tune the dynamics between regimes of fast and slow thermalisation, with the change occurring at the stochastic point (a quantum analogue of what occurs in the classical stochastic case \cite{Causer2022}). 
The same model in the fast regime ($s < 0$) was considered in Ref.~\cite{Voinea2023} as an effective model for Moore-Read states on thin cylinders.

In this paper, we have focused on the slow dynamical regime ($s > 0$), where we find an emergence of extra effective constraints that define the so-called folded model.
By means of exact diagonalisation, we have investigated the spectral statistics and relaxation dynamics for the model. 
For each of the parameter regimes studied here, our results indicate a change in the dynamical behaviour near the stochastic point. 
This is a change from a fast thermalizing dynamics to a slow thermalizing dynamics that includes metastable regimes and slow (sub-)logarithmic growth of entanglement for relevant initial states.
For the parameter that gives the most constrained effective dynamics, we also find that these initial states (which are frozen in the folded picture) appear to avoid thermalisation at all times accessible to numerics. 

This apparent violation of the ETH can be attributed to the existence of non-thermal eigenstates, throughout the spectrum of the Hamiltonian, that have a large overlap with the initial states of interest.
We verified the existence of these non-thermal eigenstates for large system sizes by means of variational MPS, even when far from the large interaction limit. 
These eigenstates are non-thermal in the sense they obey area laws and expectation values of local observables are far from those of thermal equilibrium.

Our work adds to the collective understanding of phenomena which can lead to the violation of the ETH. The key insight for understanding the observations here is the existence of the more-constrained folded dynamics, as has recently been done to understand slow heterogeneous dynamics of other kinetically constrained models \cite{Zadnik2023}.

\begin{acknowledgments}
    We are grateful to Lenart Zadnik for illuminating discussions.
    We acknowledge financial support from EPSRC Grant EP/V031201/1.
    LC was supported by an EPSRC Doctoral prize from the University of Nottingham.
    MCB was partially supported  by  the DFG (German Research Foundation) under Germany's Excellence Strategy -- EXC-2111 -- 390814868,  and Research Unit FOR 5522 (grant nr. 499180199).
    We acknowledge access to the University of Nottingham Augusta HPC service. 

\end{acknowledgments}

\appendix 

\section{Ground state phase transition} \label{appendix: ground-stae}

\begin{figure}[t]
    \centering
    \includegraphics[width=0.8\linewidth]{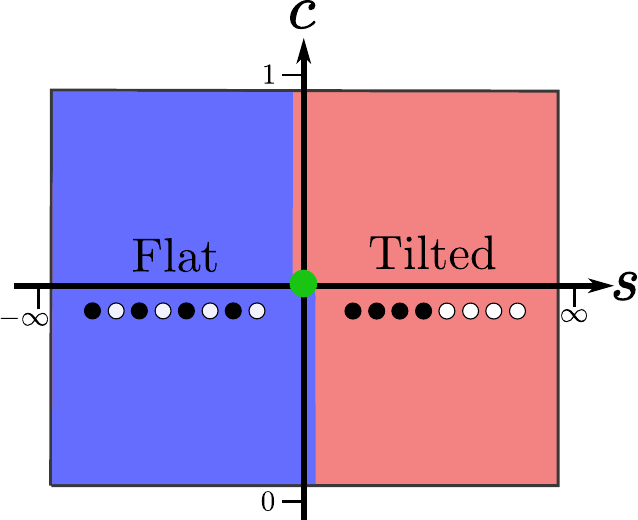}
    \caption{\textbf{Ground state phase transition.}
    Known phase diagram for the ground state of the Fredkin chain \er{H}, see e.g.\ Ref.~\cite{Causer2022}.
    The blue shaded region for $s<0$ and $s = 0$, $c < 1/2$ is the flat phase and exhibits AFM behaviour. The red shaded region for $s > 0$ and $s = 0$, $c > 1/2$ is the tilted phase and is exponentially localised with all up spins (particles) to the left.
    The green dot shows the critical point at $c = 1/2$ and $s = 0$.
    This state is maximally entropic within $\mathcal{D}$, where all possible configurations appear with equal weight in the ground state.
    }
    \label{fig: phase_diagram}
\end{figure}

The ground state of the Fredkin model in the subspace $\mathcal{D}$ has previously been shown to exhibit a rich phase diagram in the $(s, c)$ parameter space \cite{Sugino2018, Causer2022}.
The first phase corresponds to an antiferromagnet (AFM) where neighbouring spins tend to anti-align.
In analogy with dimer lattice coverings, we name this phase the {\it flat phase} \cite{Papanikolaou2007, Castelnovo2007}, as the corresponding height field encloses a small area. This exists for both  $(s<0, c)$, and $(s=0, c < 1/2)$, shown by the blue shaded region in Fig.~\ref{fig: phase_diagram}. In this phase, the ground state obeys an area law, with the von Neumann entanglement entropy for a symmetric bipartition scaling as a constant for large $N$ \cite{Sugino2018}. The second phase is a {\it localised} phase where particles are localised towards the left edge of the system and holes towards the right edge \cite{Causer2022}. We call this the {\it tilted phase}
(as the height field gives a large area), and exists for both $(s>0, c)$, and $(s = 0, c > 1/2)$, shown by the red shaded region in Fig.~\ref{fig: phase_diagram}. As for the AFM phase, the ground state also obeys an area law for its bipartite entanglement. The Hamiltonian \er{H} has a critical point at $(s = 0, c = 1/2)$ \cite{Movassagh2018, MorralYepes2023}, shown by the green dot in Fig.~\ref{fig: phase_diagram}. At this point, the ground state is an equal superposition of all possible configurations in $\mathcal{D}$ \cite{Salberger2016}, and its bipartite entanglement entropy is known to scale logarithmically in system size  \cite{Salberger2017, Udagawa2017}. Away from this critical point, the transition between the flat and tilted states is of first-order \cite{Causer2022}.

\section{Perturbation theory} \label{appendix: perturbation}
For concreteness, and because of their relevance later, we will focus on the product states $\ket{P_{j}}$ and $c > 1/2$.
Notice that $\ket{P_{j}}$ are frozen configurations in the folded model and thus are eigenstates of $\hat{H}_{c, s}^{\rm eff}$.
In what follows we will assume there are no degeneracies with the eigenstate $\ket{P_{j}}$ in the folded model.
However, in practice this should be considered carefully to ensure this is the case for the given order of perturbation theory.

The action of $\delta\hat{T}$ will sparsely connect the fragmented sectors of $\hat{H}_{c, s}^{\rm eff}$.
For first order perturbation theory, one only needs to find the fragmented sectors connected to $\ket{P_{j}}$ through the off-resonant transitions which are allowed in the original model, but not allowed in the folded model.
We numerically observe that this is at most five sectors for the states $\ket{P_{j}}$; 
 notice that $\delta \hat{T}$ only acts at the boundaries of the partitions in \er{Pj}, and so we expect the number of connecting sectors to be constant with $N$.
Furthermore, we observe that the dimensionality of each of the connecting sectors grows at most linearly in $N$, which is in contrast to the average exponential behaviour observed in Fig.~\ref{fig: fragmentation}.
Let us label the eigenstates of $\hat{H}_{c, s}^{\rm eff}$ within the subspace spanned by these five sectors, $\mathcal{M}$, by $\ket{\tilde{E}_{m}}$, with energies $\tilde{E}_{m}$.
Then the first order corrections (up to normalization) to $\ket{P_{j}}$ goes as 
\beq
    \ket{\tilde{P}_{j}^{(1)}} = \ket{P_{j}} + e^{-s} \sum_{E_{m} \in \mathcal{M}} \ket{\tilde{E}_{m}} \frac{\braket{\tilde{E}_{m} | \delta \hat{T} | P_{j}}} {E_{j} - \tilde{E}_{m}},
    \label{scar_pt_1}
\eeq
where $E_{j} = \braket{P_{j} | \hat{V} | P_{j}}$.
As the dimension of $\mathcal{M}$ is only linear in $N$, we are able to calculate \er{scar_pt_1} for system sizes up to $N \sim O(100)$.

To find the second corrections, we must now also consider the additional fragmented sectors which are connected to $\mathcal{M}$ through $\delta\hat{T}$.
We denote the subspace spanned by $\mathcal{M}$ and these additional fragmented sectors by $\mathcal{K}$. The eigenstates of $\hat{H}_{c, s}^{\rm eff}$ within $\mathcal{K}$ are again denoted by $E_{m}, E_{k}$.
The second order corrections then go as 
\begin{multline}
    \ket{\tilde{P}_{j}^{(2)}} = \ket{P_{j}} + e^{-s} \sum_{E_{m} \in \mathcal{M}} \ket{\tilde{E}_{m}} \frac{\braket{\tilde{E}_{m} | \delta \hat{T} | P_{j}}} {E_{j} - \tilde{E}_{m}},
    \\
    + e^{-2s} \sum_{E_{m} \in \mathcal{K}}\sum_{E_{k} \neq E_{m} \in \mathcal{K}} \ket{\tilde{E}_{m}} \frac{\braket{\tilde{E}_{m} | \delta \hat{T} | \tilde{E}_{k}} \braket{\tilde{E}_{k} | \delta \hat{T} | P_{j}}} {(E_{j} - \tilde{E}_{m})(E_{j} - \tilde{E}_{k})}
    \\
    -\frac{1}{2} e^{-2s} \sum_{E_{m} \in \mathcal{M}} \ket{P_{j}} \frac{|\braket{\tilde{E}_{m} | \delta \hat{T} | P_{j}}|^2}{(E_{j} - \tilde{E}_{m})^2}
    \label{scar_pt_2}
\end{multline}
The dimension of the subspace $\mathcal{K}$ is substantially larger than $\mathcal{M}$ and does not allow us to use it for such large system sizes. Nevertheless, we can use it for system sizes which can be achieved by ED to verify the eigenstates can be retrieved perturbatively. 

\section{Variational matrix product states for non-thermal eigenstates} \label{appendix: vmps}
\begin{figure}[t]
    \centering
    \includegraphics[width=\linewidth]{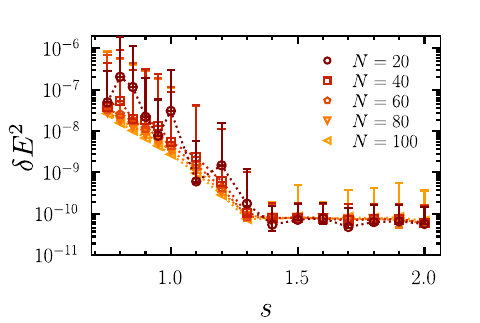}
    \caption{ 
        The energy variance of the non-thermal eigenstates estimated using vMPS for $c = 0.7$ and $N = 20, \dots, 100$.
        Each data point shows the mean energy variance over all non-thermal eigenstates $j = 1, \dots, N/2$, and the error bars show the standard deviation.
        The maximal bond dimension used was $D = 64$.
    }
    \label{fig: vmps_variance}
\end{figure}

To approximate the non-thermal eigenstates for large system sizes and some given value of $s$, we employ variational MPS.
As we are targeting eigenstates throughout the energy spectrum of the Hamiltonian, we cannot use algorithms such as DMRG which are typically used to find low-lying eigenstates.
We instead choose to minimise the energy variance \er{cost}.
Since this choice of cost function is quartic in the tensors of the MPS, we use gradient decent to minimise \er{cost} as described in Ref.~\cite{Banuls2020}.

The cost function \er{cost} is not enough to target different non-thermal eigenstates.
In Ref.~\cite{Banuls2020}, the search was steered towards states with a fixed value of the mean energy by adding a Lagrange multiplier term to the cost function, of the form $(\braket{\psi|\hat{H}|\psi}-E_{\mathrm{target}})^2$. Here, instead, we want to approximate the non-thermal eigenstate that is closest to a given $P_j$, in a similar spirit to the X-DMRG algorithm for excited states in the MBL regime in Ref.~\cite{Khemani2016xdmrg}.
To ensure we target the desired non-thermal eigenstate, we use a strategy which anneals $s$ from large-to-small.
Suppose for some given system size $N$, we wish to target the non-thermal eigenstate $\ket{{\mathcal S}_{j}}$.
In the limit $s\to\infty$, the non-thermal eigenstate will be the product state, $\lim_{s\to\infty} \ket{{\mathcal S}_{j}} = \ket{P_{j}}$.
We choose this as our initial guess for the method.
We then optimise over the sequence of MPSs
\beq
    \ket{\mathcal{\mathcal{S}}_{j}^{s_{0} = \infty}} = \ket{P_{j}} \to \ket{\tilde{\mathcal{S}}_{j}^{{s_{1}}}} \to \cdots \to \ket{\tilde{\mathcal{S}}_{j}^{{s_{T}}}},
\eeq
where $s_{0} > s_{1} > \cdots > s_{T}$ and $s_{T}$ is the target value of $s$, and $\ket{\tilde{\mathcal{S}}_{j}^{{s}}}$ is our MPS approximation of the true scarred state, $\ket{\mathcal{S}_{j}^{s}}$.

To optimise for some $s_{i}$, we propose a cost function which will aim to minimise the energy variance while maximising the overlap with the solution from the previous $s$,
\begin{multline}
    C(\ket{\tilde{\mathcal{S}}_{j}^{s_{i}}}) = \frac{\braket{\tilde{\mathcal{S}}_{j}^{s_{i}} | \hat{H}^{2} | \tilde{\mathcal{S}}_{j}^{s_{i}}}} {\braket{\tilde{\mathcal{S}}_{j}^{s_{i}} | \tilde{\mathcal{S}}_{j}^{s_{i}}}} - \frac{\braket{\tilde{\mathcal{S}}_{j}^{s_{i}} | \hat{H} | \tilde{\mathcal{S}}_{j}^{s_{i}}}^2}{\braket{\tilde{\mathcal{S}}_{j}^{s_{i}} | \tilde{\mathcal{S}}_{j}^{s_{i}}}^2} 
    \\
    - \lambda \frac{| \braket{\tilde{\mathcal{S}}_{j}^{s_{i}} | \tilde{\mathcal{S}}_{j}^{s_{i-1}}} |^2} {\braket{\tilde{\mathcal{S}}_{j}^{s_{i}} | \tilde{\mathcal{S}}_{j}^{s_{i}}}},
\end{multline}
where $\lambda > 0$ is a Lagrange multiplier.
In practice, we use a routine which slowly reduces the value of $\lambda$ to ensure convergence to an eigenstate which closely resembles the solution for the previous $s_{i-1}$.
Furthermore, we also aim to keep the bond dimension $D$ low to encourage the optimisation to find an eigenstate with small entanglement entropy.
However, we gradually increase it when required to also ensure the optimisation can find a solution with small energy variance.

The results of the method for $c = 0.7$ and $N = 20, \dots, 100$ are shown in Fig.~\ref{fig: vmps_variance}.
The plot shows the average energy variance over all scarred states $j = 1, \dots, N/2$, and the error bars show the standard deviation over all states.
We use a maximal bond dimension of $D = 64$, and terminate the optimisations when $\delta E^{2} < 10^{-10}$ or the maximal bond dimension is exceeded.
We are able to find good solutions for $s \gtrsim 0.75$.

%

\end{document}